\documentstyle[seceq,epsf,twoside]{ptptex}
\notypesetlogo  

\title{Production of 3$\pi^0$ and $\eta 2\pi^0$ from $\pi^-p$ Collision \\
in GAMS\thanks{GAMS Collaboration: D. Alde, F.G. Binon, C. Bricman, S.V. Donskov, J. Dufournaud, P. Duteil, M. Gouanere, A.V. Inyakin, V.A. Kachanov, D.B. Kakauridge, G.V. Khaustov, E.A. Knapp, A.V. Kulik, C.D. Lac, J.P. Lagnaux, A.A. Lednev, Yu.V. Mikhailov, Th. Mouthuy, V.F. Obraztsov, J.P. Peigneux, A. Possoz, Yu.D. Prokoshkin, Yu.V. Rodnov, S.A. Sadovsky, P.M. Shagin, D. Silou, A.V. Singovsky, J.P. Stroot and V.P. Sugonyaev. }
 Experiment}

\author{%
Masaaki {\sc Kobayashi}$^{1)}$, T. {\sc Tsuru}$^{1)}$, K. {\sc
Takamatsu}$^{1)}$, S. {\sc Ishida}$^{2)}$, \\
T. {\sc Komada}$^{2)}$, A. {\sc Wakabayashi}$^{2)}$ and M. {\sc Ishida}$^{3)}$ 
}
\inst{%
$^{1)}$ KEK, High Energy Accelerator Research Organization, \\ 
Tsukuba, 305-0801, Japan \\
$^{2)}$ Atomic Energy Research Institute, Nihon University, \\
Tokyo 101-8308, Japan \\
$^{3)}$Department of Physics, Tokyo Institute of Technology, \\
Tokyo 152-8551, Japan
}
\abst{%
The data on the $\pi^- p \to M^0 n$, with $M^0 = \pi^0\pi^0\pi^0$ or
$\pi^0 \pi^0 \eta$ obtained in the GAMS experiment may be useful to
study the $\sigma(400 \sim 700)$ and $a_1^\chi (\sim 1000)$, which can be taken as chiral partners of $\pi$ and $\rho$, respectively.  A preliminary analysis for $3\pi^0$ invariant mass spectra gives a support for the assumed $a_1^\chi$ with a mass of $\sim 930 MeV$ and $\Gamma \sim 170 MeV$.
}

\begin{document}
\maketitle

\setcounter{tocdepth}{4}

\vspace{-1.em}
\section{Introduction}
\vspace{-0.5em}

GAMS (NA12) experiment \cite{rf1} was carried out in 1984 at CERN SPS for the charge exchange reaction 
\vspace{0.4em} \\
\hspace*{2em}$\pi^- (100 GeV/c)$~+~$p$(Liquid $H_2$ target) ~~~~ $\to$~~~~$\gamma' s ~+~ n$  \hspace{6em} (1) 
\vspace{0.4em} \\
for different final states decaying to $\gamma$-rays.  Many papers have been published on  \\
\hspace*{1em} $ 4 \gamma' s $\cite{rf1,rf2,rf3}: ($\eta \pi^0$ ) system, rare decay of $\eta, \rho(1405) \to \eta \pi^0,G(1590) \to \eta\eta,\eta\eta'$ . \\
\hspace*{1em} $ 5 \gamma' s $\cite{rf4}: ($\omega \pi^0$) system, $\rho(2150) \to \omega \pi^0$, etc. \\
\hspace*{1em} $ 6 \gamma' s $\cite{rf5,rf6}: $\eta \pi^0 \pi^0$ for searching exotic states \\
but not much on $ 6 \gamma' s $ (especially $\pi^0\pi^0\pi^0$).

Recently, a new meson classification scheme incorporating an approximate chiral symmetry 
has been worked out by the two of the present authors and the collaborator\cite{rf7}.

 This idea leads to an explanation of the $\sigma$ meson \cite{rf8} as a chiral partner of $\pi$, and predicts many new mesons.  Especially, the following two states \\
\hspace*{3em} $\sigma(400 \sim 700)  ,~ I^G(J^{PC}) = 0^+(0^{++}),~ ^1\hspace{-0.3em}S_0$, chiral partner of $\pi(139)$, \\
\hspace*{3em} $a_1^\chi (\sim 1000) \hspace{0.9em} ,~ I^G(J^{PC}) = 1^-(1^{++}),~ ^3\hspace{-0.3em}S_1 $, chiral partner of $\rho(770)$, \\
have attracted much interest.¡¡¡¡The reactions with the final $6\gamma$
states of $\pi^0\pi^0\pi^0$ or $\eta \pi^0\pi^0$ may be useful to study
these states.  The reaction of Eq.(1) can be taken as a two step one
(see Fig. 1) consisting of (i) $\pi^-$changing to $M^0$ by the exchange
of a particle $E^-$ and (ii) decay of $M^0$ to $\pi^0\pi^0\pi^0$ or
$\eta\pi^0\pi^0$.  The exchange vertex should conserve G-parity, while
the decay vertex may break it.  
Considering that only the process of $E$ with mass
$\stackrel{<}{\scriptstyle \sim}$ 1 GeV is effective, there are two
cases of $E^-(\rho^-,\ a_0^-)$.
Then, as shown in Table I, $a_1^\chi (\sim 1000)$ should be allowed to
decay to $\pi^0\pi^0\pi^0$ but not to $\eta \pi^0 \pi^0$, in analogy to
the decay of $a_1^N$ (normal $a_1$), which has the same quantum number
as $a_1^\chi $.  $\sigma(400 \sim 700)$ decay to $2\pi^0$\cite{rf8} could be studied in both final states of $\pi^0\pi^0\pi^0$ and $\eta \pi^0 \pi^0 $.  With such an expectation, we are digging up our old data for two final states of $\pi^0\pi^0\pi^0$ and $\eta \pi^0 \pi^0 $, since the data quality should be excellent. 

\vspace{-0.5em}
\section{
Experiment and Data reduction for 6$\gamma$ states}
\vspace{-0.5em}

A $\pi^-$ beam at 100 GeV/c ( $s^{1/2} \sim 14 GeV$) was injected to a liquid $H_2$ target (6cm in diameter and 60 cm in length).  The detector system is schematically sketched in Fig. 2.  Positions and energies of all $\gamma$-rays emitted in the forward direction were measured with GAMS 4000\cite{rf2}, an array of 4096 Pb-glass modules placed at 15 m from the target.  Each lead glass module had a size of $3.8 \times 3.8 \times 45$ $cm^3$.  Two modules on 
\vspace{-1em}
\begin{figure}[hbpt]
  \epsfysize=2.5 cm
 \centerline{\epsffile{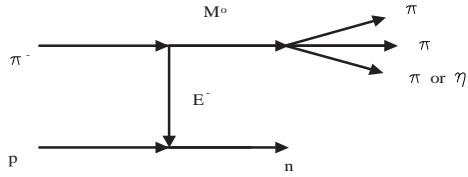}}
 \caption{ Diagram of charge exchange reaction of Eq. (1) as two step processes.}
  \label{fig:1}
\end{figure}
\vspace{-2.em}
\begin{table}
\caption{List of allowed $E^-$ and $M^0$ (see Fig. 1)
}
\begin{center}
\begin{tabular}{| l | l | l | l | l | l |}
\hline 
 & $E^-$ & $M^0$ with $J^{PC} = 0^{-+}$ & $1^{++}$ & $2^{-+}$ \\ \hline
$M^0 \to \pi^0 \pi^0 \pi^0$  & $E^- = \rho^-(770),I^G = 1^{+}$   &           & $a_1^N(1260)$       & $\pi_2(1670)$  \\
$(I^G = 1^{-})$           & G conserved                     &      
     & $a_1^\chi (\sim 1000)$  &              \\ \cline{2-5}
                         & $E^- = a_0^-(980),I^G = 1^{-}$  & $\eta(549)$  &     &  \\ 
                         & G broken                        & $\eta'(958)$ & & \\ \hline

$M^0 \to \pi^0 \pi^0 \eta$  & $E^- = a_0^-(980) $             & $\eta'(958)$ & $f_1(1285)$      &  \\
$(I^G = 0^{+})$           & G conserved                     & $\eta(1295)$ & $f_1(1420)$      &  \\
                          &                                 & $\eta(1440)$ &                  &  \\ \hline
\end{tabular}
\end{center}
\end{table}
\vspace{-0.5em}
the beam line are missing to allow the non-interacting $\pi^-$ beam pass through.  The longitudinal vertex was determined within 4 cm in $\sigma$ by measuring the Cherenkov light of the incident $\pi^-$ in the liquid $H_2$.  The transverse vertex was obtained from the fit of $\pi^0 \to 2 \gamma$ for all the existing $\pi^0$'s.  The excited baryons in the final states could be removed by detecting the deposited energy in the guard system (GS), which surrounded the target and consisted of plastic scintillators and Pb-glass.  Data analysis was carried out according to the standard scheme (see Ref.6);  Starting from raw data MT, DST and super-DST for physics analysis (called DSTA) were produced.  Experimental and MC events were treated in parallel by using essentially the same analysis program.  Fitting of events was carried out by minimizing  
\vspace{0.4em} \\
\hspace*{1em} $\chi^2 = \chi_0^2 -2\lambda_n (M_n^2-m_n^2) -2\sum \lambda_{ij}(M_{ij}^2-m_{ij}^2)$ with 
\vspace{0.4em} \\
\hspace*{1em} $\chi_0^2 = \sum_{I=1,6}
{ \{[(x_i-\underline{x_i})/\sigma(x_i)]^2 +
[(y_i-\underline{y_i})/\sigma(y_i)]^2 +
[(E_i-\underline{E_i})/\sigma(E_i)]^2} \}, $ \hspace{1em} (2)
\vspace{0.4em} \\
%
where $x_i$, $y_i$ and $E_i$ are the coordinates and the energy, respectively, of the i-th photon 
\vspace{-1em}
\begin{table}[bthp]
\caption{Statistics of the present data sample.
}
\begin{center}
\begin{tabular}{| l |}
\hline 
(1) \# of Events in DST with $\ge$6 $\gamma$'s  \hspace{5em}     1,088K  ~ events \\ 
(2) Vertex inside LH2 target                    \hspace{9.2em}    664K  \\
(3) Reject $\gamma$'s close to central hole     \hspace{6.65em}   205K  \\
(4) 6$\gamma$ events                            \hspace{16em}     117K  \\
(5) Chi-square fit to each final states (4C-fit) \hspace{2.5em}    95K           events in total \\
\hspace*{3.5em} $\pi^0 \pi^0 \pi^0$             \hspace{15.2em}    40,540 events \\
\hspace*{3.5em} $\pi^0 \pi^0 \eta$              \hspace{15.7em}    27,800  \\
\hspace*{3.5em} $\pi^0 \eta  \eta$(10,060), $\pi^0 \eta  \eta'$(6,200), $\pi^0 \pi^0 \eta'$(4,520), $\eta \eta \eta$(2,120),   \\
\hspace*{3.5em} $\eta \eta \eta'$(1,940) , $\pi^0 \eta' \eta'$(1,360) , $\eta \eta' \eta'$(580) , $\eta' \eta' \eta'$(50)  \\\hline
\end{tabular}
\end{center}
\end{table}
measured with the GAMS detector, and $\underline{x_i}$, $\underline{y_i}$ and $\underline{E_i}$ are the corresponding fitting parameters.  $\sigma(x_i)$ and $\sigma(y_i)$ were taken to be 1 cm/($E_i$ in $GeV$)$^{1/2}$.  $\lambda_{n}$ and $\lambda_{ij}$'s are also fitting parameters.  $M_n(M_{ij})$ is the neutron ($\pi^0$ or $\eta$ or $\eta'$ meson) mass calculated from the fitted variables and $m_n(m_{ij})$ is the corresponding constants given by PDG.  The degree of freedom, four, is equal to the number of the mass constraints.  We required for the accepted events (i) $\chi^2\le$12 and (ii) longitudinal vertex $=~-25\sim+25~cm$ from the target center.  The statistics of the accepted events \cite{rf6} is given in Table II.  If we require the total energy deposit (GSSUM) in GS$\le$150 MeV (noise being less than 50 MeV in each module) for neutron as recoil baryon, the statistics is reduced by 
half.  This cut is switched off if we allow contamination of excited nucleons.

\vspace{-0.5em}
\section{Preliminary result on $a_1^\chi$ and $\sigma$ for the $ 3 \pi^0 $ final state}
\vspace{-0.5em}

The $3\pi^0$ mass spectrum (Fig.3a) has a broad $\pi_2(1670)$ and a sharp $\eta(548)$ peaks.  A small but clear X(2050) peak (bump) has not yet been studied deeply.  Another peak is seen at around 950 MeV, which cannot be explained by the sharp $\eta'(958)$.

The fitting of the $3\pi^0$ mass spectrum after acceptance correction was carried out in the variant mass and width (VMW) method \cite{rf9}.  The amplitude $M$ is given by
\vspace{0.4em} \\
\hspace*{3em}$  M (s) = \sum_j r_j exp(i \theta_j) \Delta_j$ ~~with~~ $\Delta_j = m_j \Gamma_j/(s-m_j^2+im_j \Gamma_j),$ \hspace{2em}    
\vspace{-1.3em}
\begin{flushright}
(3) ~~~~
\vspace{0.4em}
\end{flushright}
%
where $r_j$ and $\theta_j$ are the amplitude and phase of the j-th
resonance, respectively.  We took the three mesons, $M^0=a_1^\chi ,\
a_1^N$ and $\pi_2(1670)$, into the amplitude which lead to the $3\pi^0$
states
as: 
(i) $a_1^\chi \to \sigma \pi^0$ , 
(ii)$a_1^N \to \sigma \pi^0, f_0(980) \pi^0$, and 
(iii) $\pi_2 \to \sigma \pi^0, f_0(980) \pi^0 , f_0^*(1270,1285) \pi^0$, 
with $\sigma, f_0, f_0^*$ all decaying to $2\pi^0$.  $\eta'(958)$ was
not included because of the absence of sharp peak in the spectrum and
the too small BR to $3\pi^0 (0.15\%)$.  Dividing the $-t=0 \sim 0.7$
region into seven at a step of 0.1, the mass and width of $\pi_2$ in
each ($-t$) region were first determined from the fit of $\pi_2$ peak in
$1.5 \sim 1.8 GeV$ .  Fixing the $\pi_2$ parameters, we then fitted all
the seven ($-t$) regions using the common parameters (mass and width)
for $a_1^N$ and $a_1^\chi$.  As seen\footnote{
In Fig. 4 only the four (out of seven) figures are shown.
} in Fig. 4, the fit with $a_1^N$ , $a_1^\chi$ and $\pi_2$ gives much better $\chi^2$/DF (456/248=1.84) than that with only $a_1^N$ and $\pi_2$ (852/264=3.23).  
The fit gives\footnote{
We also made a preliminary analysis including $\eta'(958)$ and found that its contribution to the $a_1^\chi$ peak is less than 10 \% 
and the values of mass and width are changed scarcely.}
 a mass of $931 \pm 3 MeV$ and $\Gamma$ of $166 \pm 6 MeV$ for
 $a_1^\chi$, 
while a mass of $1100 MeV$ and $\Gamma$ of $592 MeV$ for $a_1^N$. 
The evidence of $\sigma$ was searched for in the $2\pi^0$ invariant mass spectra for different $3\pi^0$ mass regions.  The $2\pi^0$ mass spectrum has a large $f_2(1270)$ peak, a small $K_s$(498) peak which may come from $\pi^- p \to \Lambda K \pi^0$, etc., and the threshold structure due to the phase space of $\eta \to 3\pi^0$ (see Figs.3b and 3c).  The significance of $\sigma$ is not very clear and is still under study.

\vspace{-0.5em}
\section{$ \pi^0 \pi^0 \eta $ final state}
\vspace{-0.5em}

Search for $\sigma$ in the $\pi^0 \pi^0$ sub-system is underway.  
We will mention general features seen in this channel (see \cite{rf6} for details); 
3-clear peaks are seen on a broad peak in the $ \pi^0 \pi^0 \eta $  mass spectrum; 
$\eta'$(958), $f_1$(1285)(more dominant than $\eta$(1295) since the moment $T_{LM}$ with $L \geq 2$ is large)$ \to a_0(980) \pi^0$, and X(1440) ($\eta$(1440) or $f_1$(1420))$\to a_0(980) \pi^0 $.  The production cross section ratio after acceptance correction is 
\vspace{0.4em} \\
\hspace*{1em} $ \sigma(\pi^- p \to \eta'(958)n){\bf B}(\eta' \to \pi^0 \pi^0 \eta) / \sigma(\pi^- p \to f_1(1285)n){\bf B}(f_1 \to \pi^0 \pi^0 \eta)$  
\vspace{0.4em} \\
\hspace*{1em} $ / \sigma(\pi^- p \to X(1440)n){\bf B}(X \to \pi^0 \pi^0 \eta ) = 100 / (29.7 \pm 1.7) / (13.5 \pm 1.4). $
\vspace{0.4em} \\
The broad peak around 1600 $MeV$ should contain significant contributions of meson peaks, 
since it cannot be explained by the acceptance-corrected phase space.  
The $\pi^0 \pi^0$ invariant mass spectrum consists of 
(i) a dominant peak of $f_2$(1270), 
(ii) a small $K_s$(498) peak, 
(iii) a broad peak around 700 $MeV/c^2$, which may contain $\sigma$ meson, and 
(iv) a threshold structure due to the decay of $\eta$'(958).  
The $\pi^0 \eta$ invariant mass spectrum contains two dominant peaks of $a_0$(980) and $a_2$(1320).

\vspace{-0.5em}
\section{Summary}
\vspace{-0.5em}
(1) We are preparing $\pi^- p \to \pi^0 \pi^0 \pi^0 n$, $\pi^0 \pi^0 \eta n$ data taken in the GAMS experiment in order to serve them for the study of \\
\hspace*{2em} $\sigma(400 \sim 700)~, I^G(J^{PC}) = 0^+(0^{++}),~ ^1\hspace{-0.3em}S_0,$ chiral partner of $\pi(139)$, \\
\hspace*{2em} $a_1^\chi (\sim 1000)~~~~, I^G(J^{PC}) = 1^-(1^{++}),~ ^3\hspace{-0.3em}S_1,$ chiral partner of $\rho(770)$, \\
(2) A preliminary analysis of $3 \pi^0$ invariant mass spectra gives a support for the $a_1^\chi$ with a mass of $\sim 930 MeV$ and $\Gamma \sim 170 MeV$. \\
(3) Analysis of $2\pi^0$ invariant mass spectra in $\pi^- p \to \pi^0 \pi^0 \pi^0 n$ is more complicated since the background increases due to 3 different combinations of $\pi^0 \pi^0$.  Preliminary analysis is not inconsistent with the existence of $\sigma(400 \sim 700)$. \\
(4) The $\pi^0 \pi^0 \eta$ channel may be convenient in searching for $\sigma \to 2\pi^0$, since the ($\pi^0 \pi^0$) combination is unique.


\begin{figure}[hbpt]
  \epsfysize=5 cm
 \centerline{\epsffile{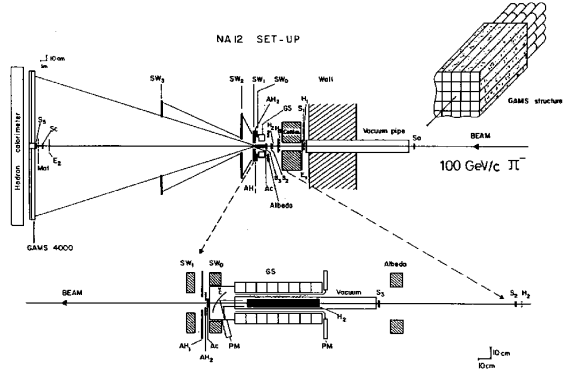}}
 \caption{ A schematic sketch of the GAMS detector.}
  \label{fig:1}
\end{figure}
\begin{figure}[hbpt]
  \epsfysize=10 cm
 \centerline{\epsffile{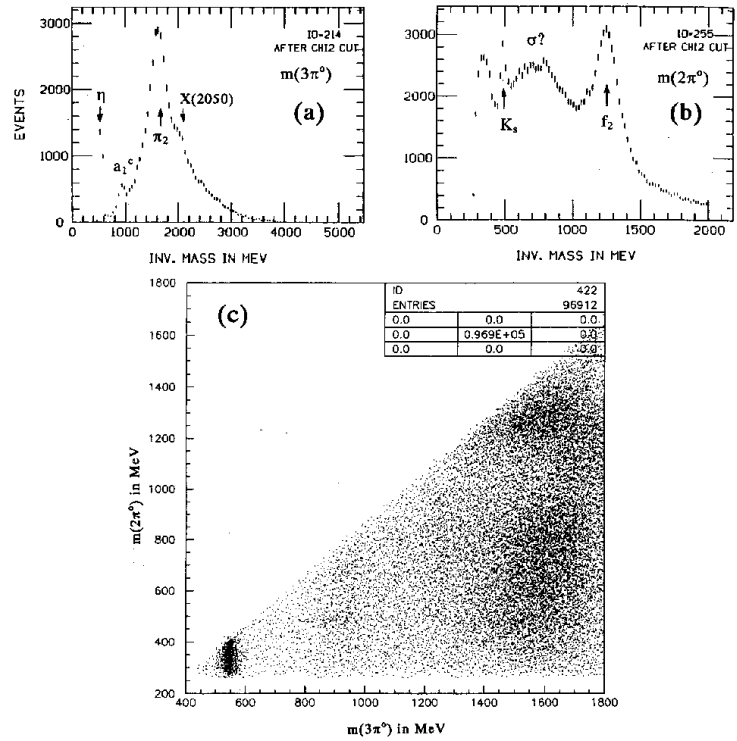}}
 \caption{ 
(a)$3\pi^0$ and 
(b)$2\pi^0$ invariant mass spectra in the $\pi^- p \to \pi^0 \pi^0 \pi^0n$ reaction for $-t \geq 0.06 GeV^2$.  
(c) A plot of $\pi^0 \pi^0 \pi^0$ in the  ($2\pi^0, 3\pi^0$) invariant mass plane for     
     -t $\le$ 0$-$1 $GeV^2$.  All figures are for $\chi^2 \le$ 12 and GSSUM$\le$150 $MeV$.
}
  \label{fig:1}
\end{figure}

\newpage
\begin{figure}[hbpt]
  \epsfysize=18 cm
 \centerline{\epsffile{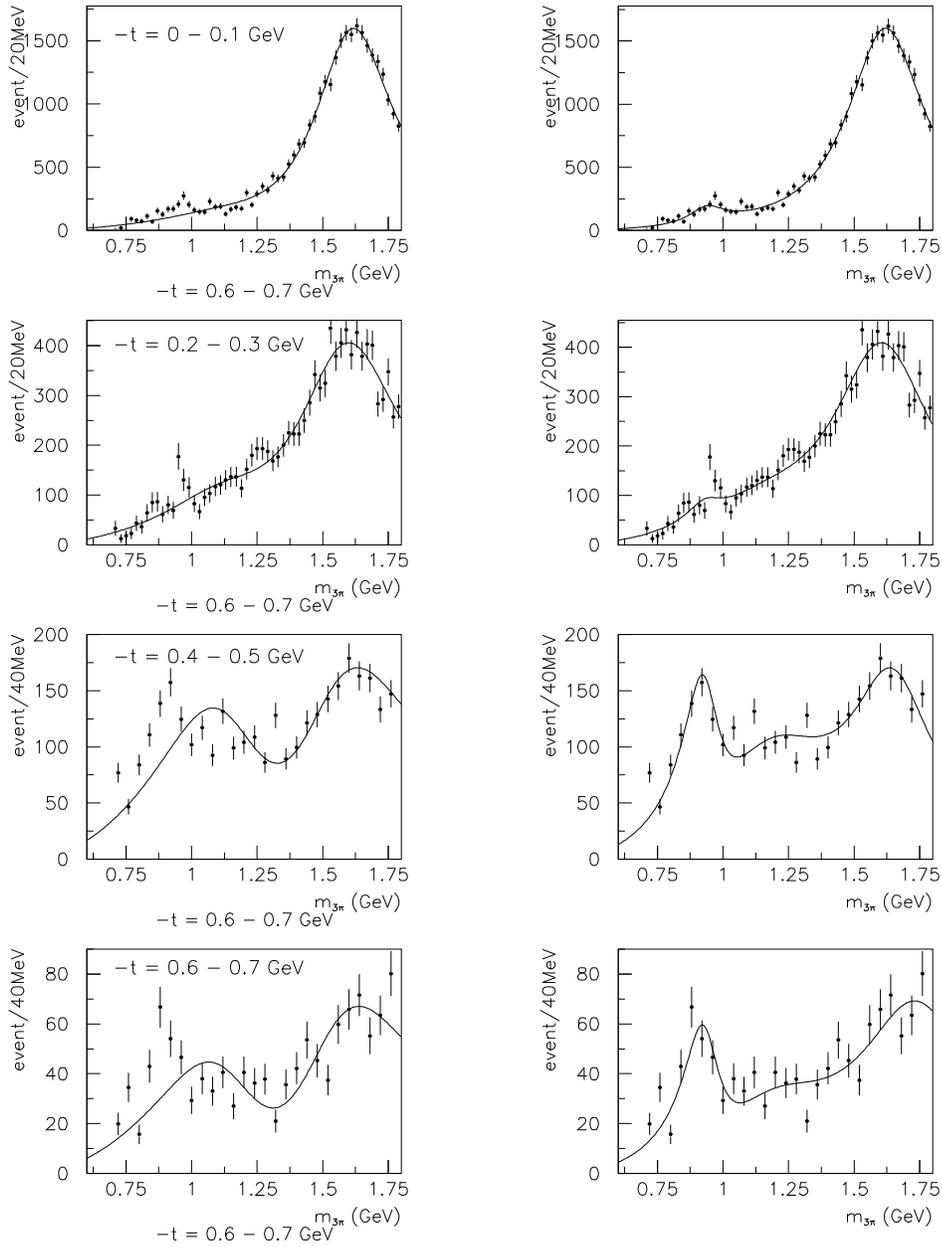}}
 \caption{
Comparison between the fits of $3\pi^0$ invariant mass spectra after acceptance 
correction with $a_1$ and $\pi_2$ (on the left), and with $a_1, a_1^\chi$ and $\pi_2$ (on the right). 
}
  \label{fig:1}
\end{figure}

\newpage

\end{document}